\documentclass{article}
\usepackage{ismir,amsmath,cite,url}
\usepackage{graphicx}
\usepackage{color}
\usepackage[nolist, nohyperlinks]{acronym}
\usepackage{multirow}
\usepackage[nottoc,numbib]{tocbibind}
\renewcommand{\footnotesize}{\scriptsize}

\title{Music mood detection based on audio and lyrics with Deep Neural Net}

\oneauthor
{\begin{tabular}{c}R\'{e}mi Delbouys \qquad Romain Hennequin \qquad Francesco Piccoli  \\ Jimena Royo-Letelier \qquad Manuel Moussallam \end{tabular}}
{Deezer,
	12 rue d'Ath\`{e}nes, 75009 Paris, France\\
	\small{\texttt{research@deezer.com}} \\
    }
\def\authorname{R\'{e}mi Delbouys, Romain Hennequin, Francesco Piccoli, Jimena Royo-Letelier, Manuel Moussallam}

\begin{document}

\maketitle

\begin{abstract}
We consider the task of multimodal music mood prediction based on the audio signal and the lyrics of a track. We reproduce the implementation of traditional feature engineering based approaches and propose a new model based on deep learning. We compare the performance of both approaches on a database containing 18,000 tracks with associated valence and arousal values and show that our approach outperforms classical models on the arousal detection task, and that both approaches perform equally on the valence prediction task. We also compare the \textit{a posteriori} fusion with fusion of modalities optimized simultaneously with each unimodal model, and observe a significant improvement of valence prediction. We release part of our database for comparison purposes.
\end{abstract}

\begin{acronym}
\acro{MFCCs}{Mel-Frequency Cepstral Coefficients}
\acro{SVM}{Support Vector Machine}
\acro{GMM}{Gaussian Mixture Model}
\acro{ConvNet}{convolutional neural network}
\acro{RNN}{recurrent neural network}
\acro{LSTM}{Long Short-Term Memory}
\acro{GRU}{Gated Recurrent Unit}
\acro{biLSTM}{Bidirectional Long Short-Term Memory}
\acro{MIR}{Music Information Retrieval}
\acro{MIREX}{Music Information Retrieval Evaluation eXchange}
\acro{TF}{Term-Frequency}
\acro{TFIDF}{Term-Frequency Inverse Document Frequency}
\acro{MSD}{Million Song Dataset}
\acro{MDS}{Multi-Dimensional Scaling}
\acro{POS}{Part-of-speech}
\end{acronym}

\section{Introduction}

\ac{MIR} has been an ever growing field of research in recent years, driven by the need to automatically process massive collections of music tracks, an important task to, for example, streaming companies. In particular, automatic music mood detection has been an active field of research in \ac{MIR} for the past twenty years. It consists of automatically determining the emotion felt when listening to a track.\footnote{We use the words emotion and mood interchangeably, as done in the literature (see \cite{kim2010music}).} In this work, we focus on the task of multimodal mood detection based on the audio signal and the lyrics of the track. We apply deep learning techniques to the problem and compare our approach to classical feature engineering-based ones on a database of 18,000 songs labeled with a continuous arousal/valence representation. This database is built on the \ac{MSD} \cite{Bertin-Mahieux2011} and the Deezer catalog.  To our knowledge this constitutes one of the biggest datasets for multimodal mood detection ever proposed.
\subsection{Related work}
Music mood studies appeared in the first half of the $20$th century, with the work of Hevner 
\cite{hevner1936experimental}. In this work, the author defines groups of emotions and studies classical music works to unveil correlations between emotions and  characteristics of the music. A first indication that music and lyrics should be jointly considered when analyzing musical mood came from a psychological study exposing independent processing of these modalities by the human brain \cite{besson1998singing}. For the past 15 years, different approaches have been developed with a wide range of datasets and features. An important fraction of them was put together by Kim et al.\ in \cite{kim2010music}. Li and Ogihara \cite{li2003detecting} used signal processing features related to timbre, pitch and rhythm. Tzanetakis et al.\ \cite{tzanetakis2007marsyas} and Peeters \cite{peetersMIREX08} also used classical audio features, such as \ac{MFCCs}, as input to a \ac{SVM}. Lyrics-based mood detection was most often based on feature engineering. For example, Yang and Lee \cite{yang2004disambiguating} resorted to a psycho-linguistic lexicon related to emotion. Argamon et al.\ \cite{argamon2003style} extracted stylistic features from text in an author detection task. Multimodal approaches were also studied several times. Laurier et al.\ \cite{laurier2008multimodal} compared prediction level and feature level fusion, referred to as late and early fusion respectively. In \cite{su2017graph},  Su et al.\ developed a sentence level fusion. An important part of the work based on feature engineering was compiled into more complete studies, among which the one from Hu and Downie \cite{hu2010improving}  is one of the most exhaustive, and compares many of the previously introduced features.

Influenced by advances in deep learning, notably in speech recognition or machine translation, new models began to emerge, based on fewer feature engineering. Regarding audio-based methods, the \ac{MIREX} competition \cite{downie2006music} has monitored the evolution of the state of the art. In this framework, Lidy et al.\ \cite{lidy2016parallel} have shown the promise of audio-based deep learning. Recently, Jeon et al.\ \cite{jeonmusic} presented the first multimodal deep learning approach using a bimodal convolutional recurrent network with a binary mood representation. However, they neither compared their work to classical approaches, nor evaluated the advantage of their mid-level fusion against simple late fusion of unimodal models. In \cite{huang2016bi}, Huang et al.\ resorted to deep Boltzmann machines to unveil early correlations between audio and lyrics, but their method was limited by the incompleteness of their dataset, which made impossible the use of temporally local layers, e.g.\ recurrent or convolutional ones.  To our knowledge, there is no clear answer as to whether feature engineering yields better results than more end-to-end systems for the multimodal task,  probably because of the lack of easily accessible large size datasets.

\subsection{Mood representation}
A variety of mood representations have been used in the literature. They either consist of monolabel tagging with either simple tags (e.g.\ in \cite{hu2010improving}), clusters of tags (e.g.\ in the \ac{MIREX} competition) or continuous representation. In this work, we resort to the latter option. Russell \cite{russell1980circumplex} defined a 2-dimensional continuous space of embedding for emotions. A point in this space represents the valence (from negative to positive mood) and arousal (from calm to energetic mood) of an emotion. This representation was used multiple times in the literature \cite{huang2016bi, trigeorgis2016adieu,wang2011music}, and presents the advantage of being satisfyingly exhaustive. It is worth noting that this representation has been validated by embedding emotions in a 2-dimensional space based on their co-occurrences in a database \cite{hu2010lyrics}.  Since we choose this representation we formulate mood estimation as a 2-dimensional regression problem based on a track's lyrics and/or audio.

\subsection{Contributions of this work}
We study end-to-end lyrics-based approaches to music mood detection and compare their performance with classical lyrics-based methods performance, and give insights on the performing architectures and networks types. We show that lyrics-based networks show promising results both in valence and arousal prediction.

We describe our bimodal deep learning model and evaluate the performance of a mid-level fusion, compared to unimodal approaches and to late fusion of unimodal predictions. We show that arousal is highly correlated to the audio source, whereas valence requires both modalities to be predicted significantly better. We also see that the latter task can be  notably improved by resorting to mid-level fusion.

Finally, we compare our model to traditional feature engineering methods and show that deep-learning-based approaches outperform classical models, when it comes to multimodal arousal detection, and we show that both systems are equally performing on valence prediction. For future comparison purposes, we also release part of our database consisting of valence/arousal labels and corresponding song identifiers.

\section{Classical feature engineering-based approaches}\label{sec:classicalapp}
We compare our model to classical approaches based on feature engineering. These methods were iteratively deepened over the years: for audio-based models, a succession of works \cite{li2003detecting, peetersMIREX08, tzanetakis2007marsyas} indicated the top performing audio features for mood detection tasks ; for lyrics-based approaches, a series of studies \cite{argamon2003style,yang2004disambiguating,hu2010lyrics} investigated a wide variety of text-based features. Finally, fusion methods were also studied multiple times \cite{laurier2008multimodal, wang2011music, hu2010improving}. Hu and Downie compiled and deepened these works in a series of papers \cite{hu2010lyrics,hu2010improving,hu2016framework}, which is the most accomplished feature-engineering-based approach of the subject. We reimplement this work and compare its performance to ours. This model consists in the choice of the optimal weighted average of the predictions of two unimodal models: an \ac{SVM} on top of \ac{MFCCs}, spectral flux, rolloff and centroid, for audio; and an \ac{SVM} on top of basic, linguistic and stylistic features (n-grams, lexicon-based features, etc.) for lyrics.  

\section{Deep learning-based approach}
We first explore unimodal deep learning models and then combine them into a multimodal network. In each case, the model simultaneously predicts valence and arousal. Inputs are subdivided in several segments for training, so that each input has the same length. Output is the average of the predictions computed by the model on several segments of the input. For the bimodal models, subdivision of audio and lyrics requires synchronization of the modalities. 

\subsection{Audio only}\label{ssec:audioapp}
\begin{figure}[htb]

\begin{minipage}[b]{0.24\linewidth}
  \centering
  \centerline{\includegraphics[width=1.9cm]{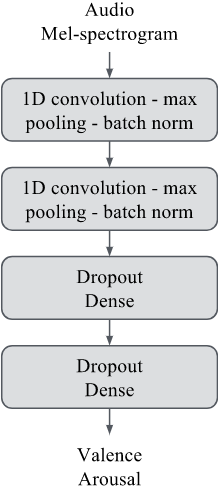}}
  \centerline{(a) Audio}\medskip
\end{minipage}
\begin{minipage}[b]{.24\linewidth}
  \centering
  \centerline{\includegraphics[width=1.9cm]{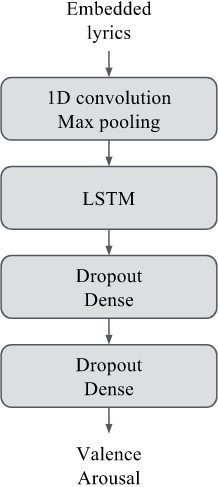}}
  \centerline{(b) Lyrics}\medskip
\end{minipage}
\hfill
\begin{minipage}[b]{0.49\linewidth}
  \centering
  \centerline{\includegraphics[width=3.9cm]{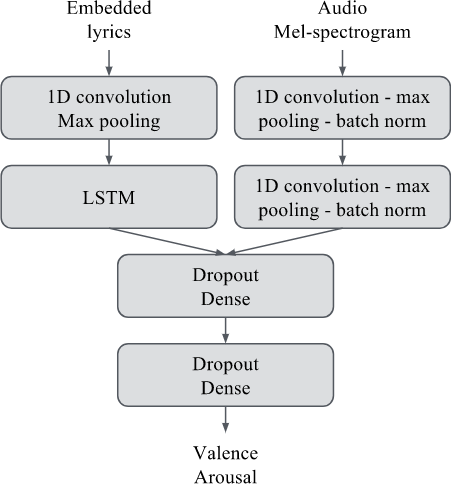}}
  \centerline{(c) Bimodal}\medskip
\end{minipage}
\caption{Architecture of unimodal and bimodal models}
\label{fig:architectures}
\end{figure}

We use a mel-spectrogram as input, which are 2-dimensional. We choose a \ac{ConvNet} \cite{lecun2010convolutional}, the architecture is shown in Fig.\ \ref{fig:architectures} (a). It is composed of two consecutive $1$-dimensional convolution layers (convolutions along the temporal dimension) with $32$ and $16$ feature maps of size $8$, stride $1$, and max pooling of size $4$ and stride $4$. We resort to batch normalization \cite{ioffe2015batch} after each convolutional layer. We use two fully connected layers as output to the network, the intermediate layer being of size $64$.

\subsection{Lyrics only}
We use a word embedding as input to the network, i.e.\ each word is embedded in a continuous space and the vectors corresponding to each word are stacked, the input being consequently 2-dimensional. We choose to resort to a word2vec \cite{mikolov2013efficient} embedding trained on 1.6 million lyrics, as first results seemed to indicate that this specialized embedding performs better than embedding pretrained on an unspecialized, albeit bigger, dataset. We compare several architectures, with recurrent and convolutional layers. One of them is shown in Fig.\ \ref{fig:architectures} (b). We also compare this approach with a simple continuous bag-of-words method that acts as a feature-free baseline. The models that were tested are described in Table \ref{table:lyrics_models}. 
\begin{table}
\centering
\resizebox{\linewidth}{!}{\begin{tabular}{c|l}
  \hline
   Model name & Description \\
  \hline
     CBOW&\begin{minipage}[t]{0.8\linewidth}
            \begin{text}
                Continuous bag-of-words: random forest on top of means of input words embedding
            \end{text}
        \end{minipage} \rule[-4ex]{0pt}{0pt} \\ 
     GRU&\begin{minipage}[t]{0.8\linewidth}
            \begin{text}
                Single \ac{GRU} \cite{cho2014properties}, size \(40\), dense layers of size \(64\) and \(2\), preceded by dropout layers of parameter \(0.5\)
            \end{text}
        \end{minipage} \rule[-4ex]{0pt}{0pt}\\ 
     LSTM&\begin{minipage}[t]{0.8\linewidth}
            \begin{text}
                Single \ac{LSTM} \cite{gers1999learning}, size \(80\), dense layers of size \(64\) and \(2\), preceded by dropout layers of parameter \(0.5\)
            \end{text}
        \end{minipage} \rule[-4ex]{0pt}{0pt}\\ 
     biLSTM&\begin{minipage}[t]{0.8\linewidth}
            \begin{text}
                Single \ac{LSTM}, size \(40\), dense layers of size \(64\) and \(2\), preceded by dropout layers of parameter \(0.5\)
            \end{text}
        \end{minipage} \rule[-4ex]{0pt}{0pt}\\ 
     2LSTMs &\begin{minipage}[t]{0.8\linewidth}
            \begin{text}
                Two \ac{LSTM} layers, of size  \(40\), dense layers of size \(64\) and \(2\), preceded by dropout layers of parameter \(0.5\)
            \end{text}
        \end{minipage} \rule[-4ex]{0pt}{0pt}\\ 
        ConvNet+LSTM &\begin{minipage}[t]{0.8\linewidth}
            \begin{text}
                Convolutional layer with \(16\) features maps of size (\(2\),\(2\)), stride $1$, max-pooling of size \(2\), stride $2$, an \ac{LSTM} layer of size \(40\) and dense layers of size \(32\) and \(2\), preceded by dropout layers of parameter \(0.5\)
            \end{text}
        \end{minipage} \rule[-4ex]{0pt}{0pt}\\ 
        2ConvNets+2LSTMs &\begin{minipage}[t]{0.8\linewidth}
            \begin{text}
               Two convolutional layers with \(16\) features maps of size (\(2\),\(2\)), stride $1$, max-pooling of size \(2\), stride $2$, two \ac{LSTM} layers of size \(40\) and dense layers of size \(32\) and \(2\), preceded by dropout layers of parameter \(0.5\)
            \end{text}
        \end{minipage} \rule[-4ex]{0pt}{0pt}\\ 
  \hline
\end{tabular}}
\caption{Description of lyrics-based models.}
\label{table:lyrics_models}
\end{table}

\subsection{Fusion}\label{ssec:midfusionapp}
For the fusion model, we reuse the unimodal architecture from which we remove the fully connected layers and concatenate the outputs of each network. On top of this concatenation, we use two fully connected layers with an intermediate vector length of size \(100\). This architecture is presented in Fig.\ \ref{fig:architectures}(c). This allows for detection of more complex correlations between modalities. We choose to compare this with a simple late fusion, which is a weighted average of the outputs of the unimodal models, the weight being grid-searched. The mid-level fusion model is referred to as \texttt{middleDL} and the late fusion model as \texttt{lateDL}.

\section{Experiment}
\subsection{Dataset}
The \ac{MSD} \cite{Bertin-Mahieux2011} is a large dataset commonly used for \ac{MIR} tasks. The tracks are associated with tags from LastFM\footnote{\url{http://www.last.fm/}}, some of which are related to mood. We apply the procedure described by Hu and Downie in \cite{hu2009lyric} to select the tags that are akin to a mood description.  We then make use of the dataset published by Warriner et al.\ \cite{warriner2013norms} which associates 14,000 English words with their embedding in Russell's valence/arousal space. We use it for embedding previously selected tags into the valence/arousal space. When several tags are associated with the same track, we retain the mean of the embedding values. Finally, we normalize the database by centering and reducing valence and arousal. It would undoubtedly be more accurate to have tracks directly labeled with valence/arousal values by humans, but no database with sufficient volume exists. An advantage of this procedure is its applicability to different mood representations, and thus to different existing databases.

The raw audio signal and lyrics are not provided in the \ac{MSD}. Only features are available, namely \ac{MFCCs} for audio, word-counts for lyrics. For this reason, we use a mapping between the \ac{MSD} and the Deezer catalog using the song metadata (song title, artist name, album title) and have then access to raw audio signals and original lyrics for a part of the songs. As a result, we collected a dataset of 18,644 annotated tracks. We note that lyrics and audio are not synchronized. Automatic synchronization being outside of the scope of this work, we resort to a simple heuristic for audio-lyrics alignment. It consists of aligning both modalities proportionally based on their respective length, i.e.\ for a certain audio segment, we extract words from the lyrics that are at the corresponding location relatively to the length of the lyrics. We release the labels, along with Deezer song identifiers, \ac{MSD} identifiers, artist and track name\footnote{\url{https://github.com/deezer/deezer_mood_detection_dataset}}. More data can be retrieved using the Deezer API\footnote{\url{https://developers.deezer.com/api}}. Unfortunately, we cannot release the lyrics and music, due to rights restrictions.

We train the models on approximately $60\%$ of the dataset, and validate their parameters with another $20\%$. Each model is then tested on the remaining $20\%$. We refer to these three sets as training, validation and test set, respectively. We split the dataset randomly, with the constraint that songs by the same artist must not appear in two different sets (since artist and moods may be correlated).

\subsection{Implementation details}
For audio, we use a mel-spectrogram as input to the network, with $40$ mel-filters and $1024$ sample-long Hann window with no overlapping, with a sampling frequency of $44.1$kHz, computed with YAAFE \cite{mathieu2010yaafe}. We use data augmentation, that was investigated for audio and proven useful in \cite{schluter2014improved}, in order to grow our dataset. First, we decide to extract \(30\) second long segments from the original track. The input of the network is consequently of size $40$*$1292$. We choose to sample seven extracts per track: we draw them uniformly from the song. We also use pitch shifting and lossy encoding, which are transformations with which emotion is invariant, and get three extra segments per original sample. In the end, we get a $28$-fold increase in the size of the training set.

For lyrics, the input word embedding was computed with gensim's implementation of word2vec \cite{rehurek_lrec} and we used \(100\)-dimensional vectors. We use data augmentation for lyrics as well by extracting seven \(50\)-word segments from each track. Consequently, the input of each neural network is of size $100$*$50$.

\subsection{Results}
\begin{table}
\centering
\begin{tabular}{c|c|c|c}
  \hline
  mode & model & valence & arousal\\
  \hline
   \multirow{2}{*}{audio} & CA & 0.118 & 0.197\\
   & ConvNet & \textbf{0.179} & \textbf{0.235}\\
  \hline
   \multirow{6}{*}{lyrics} & CA & \textbf{0.140 }& \textbf{0.032}\\
   & CBOW &0.080 & 0.031\\
   & LSTM & 0.117 & 0.027\\
   & GRU & 0.106 & 0.017\\
   & biLSTM & 0.076 & 0.017\\
   & 2LSTMs & 0.128 & 0.024\\
   & ConvNet+LSTM & 0.134 & 0.026\\
   & 2ConvNets+2LSTMs & 0.127 & 0.022\\
  \hline
   \multirow{3}{*}{bimodal} & CA & \textbf{0.219} & 0.216\\
   & LateDL & 0.194 & \textbf{0.235}\\
   & middleDL & \textbf{0.219} & \textbf{0.232}\\
  \hline
\end{tabular}
\caption{$R^2$ scores of the different tested approaches.}
\label{table:allres}
\end{table}

\begin{table*}
\centering
\resizebox{\textwidth}{!}{\begin{tabular}{c|c|ccccccccccc}
  \hline
   & coefficient* & 0.0 & 0.1 & 0.2 & 0.3 & 0.4 & 0.5 & 0.6 & 0.7 & 0.8 & 0.9 & 1.0\\
  \hline
   \multirow{2}{*}{Feature engineering approaches} & valence &  0.133  &  0.163  &  0.186  &  0.201  &  \textbf{0.211 } &  \textbf{0.211}  &  0.207  &  0.192  &  0.174  &  0.147  &  0.112\\
   & arousal & 0.034  &  0.081  &  0.121  &  0.152  &  0.178  &  0.199  &  0.211  &  0.217  &  \textbf{0.218 } &  0.212  &  0.201 \\
  \hline
   \multirow{2}{*}{Deep learning approaches} & valence & 0.118  &  0.136  &  0.152  &  0.165  &  0.175  &  0.182  &  0.186  &  \textbf{0.188}  &  0.187  &  0.183  &  0.177\\
& arousal & 0.025  &  0.065  &  0.102  &  0.135  &  0.164  &  0.19  &  0.212  &  0.231  &  0.246  &  0.257 &  \textbf{0.265}  \\
  \hline
\end{tabular}}
\caption{$R^2$ scores of the late fusion of unimodal models for classical approaches and deep learning approaches, for different values of weighting. *This coefficient is the weight of the audio prediction. The weight of the lyrics prediction is its complementary to one.}
\label{table:latefusion}
\end{table*}

We present the results and compare in particular deep learning approaches with classical ones. The results are presented in Tab.\ \ref{table:allres} and \ref{table:latefusion}. In the latter, \texttt{CA} refers to classical models  (described in Sect.\ \ref{sec:classicalapp}).

%
\begin{table}
\centering
\begin{tabular}{c|c|c|c|c}
  \hline
   \multirow{2}{*}{modalities} & \multirow{2}{*}{BWC*} & CA and DL & \multirow{2}{*}{CA} & \multirow{2}{*}{DL} \\
    &  & mean &  &  \\
  \hline
   audio & 0.7 &  \textbf{0.193} & 0.118 & 0.179 \\
   lyrics & 0.5 & \textbf{0.177} & 0.140 & 0.134 \\
    fused   &  0.5 & \textbf{0.243} & 0.219 & 0.219 \\
  \hline
\end{tabular}
\caption{$R^2$ scores of the optimal weighted mean of classical and deep learning approaches for valence prediction for different modalities. *BWC: best weighting coefficient. This coefficient is the optimal weight of the deep learning-based prediction. CA and DL respectively refers to classical approaches and deep learning methods.\vspace{-1em}}
\label{table:latefusionhdnn}
\end{table}

\textbf{Unimodal approaches.} The results of each unimodal model are given in Table \ref{table:allres}. For lyrics-based ones, we have tested several models without feature engineering. The highest performing method, on both validation and test set, is based on both recurrent and convolutional layers. In the following, we choose this model as the one to be compared with classical models.

For both unimodal models, one can see a similar trend for classical and deep learning approaches: lyrics and audio achieve relatively similar performance on valence detection, whereas audio clearly outperforms lyrics when it comes to arousal prediction. This is unsurprising, as arousal is closely related to rhythm and energy, which are essentially induced by the audio signal. On the contrary, valence is explained by both lyrics and audio, indicating that the positivity of an emotion can be conveyed through the text as well as through the melody, the harmony, the rhythm, etc. Similar observations were made by Laurier et al.\ \cite{laurier2008multimodal}, where angry and calm songs were classified significantly better by audio than by lyrics, and happy and sad songs were equally well-classified by both modalities. This is consistent with our observations, as happy and sad emotions can be characterized by high and low valence, and angry and calm emotions by high and low arousal.

When looking more closely at the results, one can observe that deep learning approaches are much higher performing than classical ones when it comes to prediction based on audio. On the contrary, classical lyrics-based models are higher performing than our deep learning model, in particular when it comes to valence detection, which is the most informative task for the study on lyrics only (as stated above). The reason can be that classical systems resort to several emotion related lexicons designed by psychological studies. On the contrary, classical audio feature engineering for mood detection does not make use of such external resources curated by experts. 

\textbf{Late fusion analysis.} As stated earlier, the late fusion consists of a simple optimal weighted average between the prediction of both unimodal models. We resort to a grid-search on the value of the weighting between 0 and 1. The result for the reimplementation of traditional approaches and for our model is presented in Table \ref{table:latefusion}. One can observe a similar phenomenon for both classical models and ours. In both cases, the fusion of the modalities does not significantly improve arousal detection performance compared to audio-based models. It is as predicted, as we saw that audio-based models perform significantly better than lyrics-based ones. For deep learning models, using lyrics in addition to audio in a late fusion scheme leads to no improvement, so there is no gain added by using lyrics. When it comes to valence detection, both modalities are valuable: in both approaches, the top performing model is a relatively balanced average of unimodal predictions. Here also, these observations generalize to valence/arousal what was observed on the emotions happy, sad, angry and calm in \cite{laurier2008multimodal}. Indeed, based on this study, not only are lyrics and audio equally performant for predicting happy and sad songs, but they are also complementary, so that fused models can achieve notably better accuracies. However, predicting angry and calm songs is not improved when using lyrics in addition to audio. 

\textbf{Bimodal approaches comparison.} Bimodal method performances are reported in Table \ref{table:allres}. Several interesting remarks can be made based on these results. First of all, one can notice that if one compares late fusion for both approaches, arousal detection is outperformed by deep learning systems, as the corresponding unimodal approach based on audio is more performant, and we have seen that lyrics-based arousal detection is in both cases performing poorly. On the contrary, late fusion for valence detection yields better results for classical systems. In this case, the lack of performance of lyrics-based methods relying on deep learning is not compensated for by a slightly improved audio-based performance.

However, when it comes to mid-level fusion presented in paragraph \ref{ssec:midfusionapp}, there is a clear improvement for valence detection. It seems to indicate that there might be earlier correlations between both modalities, that our model is able to detect. Concerning arousal detection, the capacity of the network to unveil such correlations seems useless: we have seen that our lyrics-based model is not able to bring additional information to the audio-based model.

This performing fusion, along with more accurately predicted valence thanks to audio, is sufficient for achieving similar performance to classical approaches, without the use of any external data designed by experts. Interestingly, both models remain useful, as long as they learn complementary information. For valence detection, an optimized weighted average of the predictions of both models yields the performance presented in Table \ref{table:latefusionhdnn}. We can see a significant gain obtained for a balanced average of both predictions, indicating that both models have different applications, in particular when it comes to lyrics-based valence detection.

\section{Conclusion and future work}
We have shown that multimodal mood prediction can go without feature engineering, as deep learning-based models achieve better results than classical approaches on arousal detection, and both methods perform equally on valence detection. It seems that this gain of performance is the results of the capacity of our model to unveil and use mid-level correlations between audio and lyrics, particularly when it comes to predicting valence, as we have seen that for this task, both modalities are equally important. 

The gain of performance obtained when using this fusion instead of late fusion indicates that further work can be done for understanding correlations between both modalities, and there is no doubt that a database with synchronized lyrics and audio would be of great help to go further. Future work could also rely on a database with labels indicating the degree of ambiguity of the mood of a track, as we know that in some cases, there can be significant variability between listeners. Such databases would be particularly helpful to go further in understanding musical emotion. Temporally localized label in sufficient volume can also be of particular interest. Future work could also leverage unsupervised pretraining to deep learning models, as unlabeled data can be easier to find in high volume. We also leave it as a future work to pursue improvements of lyrics-based models, with deeper architectures or by optimizing word embeddings used as input. Studying and optimizing in detail \ac{ConvNet}s for music mood detection offers the opportunity to temporally localize zones responsible for the valence and arousal of a track, which could be of paramount importance to understand how music, lyrics and mood are correlated. Finally, by learning from feature engineering approaches, one could use external resources designed by psychological studies to improve significantly the prediction accuracy, as indicated by the complementarity of both approaches.

\section{Acknowledgments}
The authors kindly thank Geoffroy Peeters and Gabriel Meseguer Brocal for their insights as well as Matt Mould for his proof-reading. The research leading to this work benefited from the WASABI project supported by the French National Research Agency (contract ANR-16-CE23-0017-01).

\section{REFERENCES}
\label{sec:refs}

\bibliographystyle{IEEE}
\begingroup
\renewcommand{\section}[2]{}%
\bibliography{biblio}
\endgroup

\end{document}